\documentclass[a4paper]{article}
\usepackage{float}
\pdfoutput=1
\usepackage{INTERSPEECH2018}
\usepackage{here}
\usepackage{url}
\usepackage{array,etoolbox}
\usepackage{caption}
\preto\tabular{\setcounter{magicrownumbers}{0}}
\newcounter{magicrownumbers}
\def\rownumber{}

\DeclareMathOperator*{\argmax}{argmax}

\title{Student-Teacher Learning for BLSTM Mask-based Speech Enhancement}
\name{Aswin Shanmugam Subramanian, Szu-Jui Chen, Shinji Watanabe}

\address{
  Center for Language and Speech Processing, Johns Hopkins University}
\email{\{aswin, schen146, shinjiw\}@jhu.edu}

\begin{document}

\maketitle 
\begin{abstract}
Spectral mask estimation using bidirectional long short-term memory (BLSTM) neural networks has been widely used in various speech enhancement applications, and it has achieved great success when it is applied to multichannel enhancement techniques with a mask-based beamformer.
However, when these masks are used for single channel speech enhancement they severely distort the speech signal and make them unsuitable for speech recognition. 
This paper proposes a student-teacher learning paradigm for single channel speech enhancement. 
The beamformed signal from multichannel enhancement is given as input to the teacher network to obtain soft masks. 
An additional cross-entropy loss term with the soft mask target is combined with the original loss, so that the student network with single-channel input is trained to mimic the soft mask obtained with multichannel input through beamforming. 
Experiments with the CHiME-4 challenge single channel track data shows improvement in ASR performance.

\end{abstract}

\section{Introduction}
\label{sec:intro}
The presence of background noise and reverberation degrades the performance of an automatic speech recognition (ASR) system. 
The performance of noise robust ASR was greatly improved by using multiple microphones instead of just using a single microphone \cite{barker2015third,kinoshita2016summary,li2017acoustic}.
Especially, mask-based beamforming techniques have shown outstanding results in this scenario \cite{higuchi2016robust,nn-gev,erdogan2016improved}, and many of the top systems in the CHiME-4 challenge \cite{vincent2017analysis} use these techniques for speech enhancement. 
For example, \cite{nn-gev,erdogan2016improved} use a bidirectional long short-term memory (BLSTM) neural network to accurately predict speech (and noise) masks, given the noisy spectrogram.
In \cite{nn-gev}, both speech and noise masks are in turn used to calculate cross power spectral density (PSD) matrices of the noise and the target speech, which are used by the generalized eigenvalue (GEV) beamformer to get the beamforming filter \cite{warsitz2007blind}.

However, compared with the success of multichannel speech enhancement, single-channel speech enhancement is not well established, especially when the method is combined with other speech processing applications including ASR and speaker recognition \cite{hori2017multi}.
%
Single-channel speech enhancement, especially based on deep neural networks (DNNs), has been studied by many research groups \cite{lu2013speech,xu2014experimental,narayanan2013ideal,erdogan2015phase}
with promising performance improvement in terms of the speech enhancement metrics or hearing purposes.
However, we often observe a degradation in performance when we use single-channel enhancement as a preprocessing step for ASR due to the spectral distortions induced by the enhancement.
This paper focuses on the single-channel speech enhancement to overcome this issue.

Our idea is to fill out the gap between single-channel and multichannel speech enhancement by using a well known DNN technique called student-teacher learning.
We propose to train a \textit{teacher} network, which takes the beamformed signal obtained from multichannel speech enhancement as input and predicts high-quality speech masks. 
Then, a \textit{student} network, which takes the noisy signal as input, is trained to mimic the mask predicted by the teacher network. 
Conventionally student-teacher learning \cite{nips_st,jinyu_st} is used to reduce the complexity of the network. 
For example, a less complex student model with fewer parameters tries to mimic the soft targets of a more complex teacher model. 
In \cite{shinji_stl}, student-teacher training is used for self-supervised learning of an ASR acoustic model, where the student model tries to mimic the effects of multichannel speech enhancement by using only single channel inputs.
This was achieved by training the student network with noisy training data as input to mimic the soft posteriors of the teacher network trained on the enhanced speech. 
In addition, the benefits of using soft targets compared to hard targets for DNN acoustic models are also shown in \cite{shinji_stl}, which is advocated by \cite{hinton2015distilling} as knowledge distillation.

This paper follows BLSTM-mask based beamforming proposed in \cite{nn-gev}, which has two cross-entropy loss terms - one for speech mask target and the other for noise mask target.
The speech mask predicted by the model can also be used for single-channel enhancement, which this paper uses as the baseline.
We have an additional loss term based on the cross entropy with the soft mask from the teacher network as a target.
The effectiveness of the proposed method is investigated by using the single channel track of the CHiME-4 dataset \cite{chime4}. 
The CHiME-4 dataset consists of both real and simulated data and it is common to use only the simulation data to prepare the clean and noise targets.
As the proposed method uses the soft mask from the teacher network as the target, which can be obtained even for the real data, it can use both real and simulation data in the training stage. This is also an unique aspect of the proposed method.
Additionally, four different speech enhancement metrics - perceptual evaluation of speech quality (PESQ) \cite{pesq},  short-time objective intelligibility measure (STOI) \cite{stoi}, extended STOI (eSTOI) \cite{estoi} and speech distortion ratio (SDR) \cite{sdr} are used as part of the experiments to discuss the performance of speech enhancement in addition to the ASR performance.

\section{BLSTM mask-based speech enhancement}
This section first describes a mask prediction method by using a binary cross entropy loss, which is a basic component of this paper.
It also describes a mask-based beamformer to estimate beamforming filters based on the estimated masks.
\label{sec:blstm}
\subsection{Binary cross entropy loss}
\label{sec:sing}
The noise-aware training of BLSTM mask proposed in \cite{nn-gev} is explained in this section. 
The network estimates two masks: the first is the speech mask which denotes the degree of how dominant the speech component is in each time-frequency bin, while the second is the noise mask which denotes that of the noise component \cite{erdogan2016multi}. 
The ideal binary speech mask target $\textrm{IBM}_{\text{X}}(t,b) \in \{0, 1\}$ at frame $t$ in frequency bin $b$ is defined based on the signal-to-noise (SNR) ratio with thresholding as:
\begin{equation}
\textrm{IBM}_{\text{X}}(t,b) =
\begin{cases}
1, & \text{if } \frac{\Vert x (t, b) \Vert}{\Vert n (t, b) \Vert} > threshold \\
0, & \text{otherwise }
\end{cases}
,
\end{equation}
where, $\Vert x (t, b) \Vert \in \mathbb{R} _{\geq 0}$ is the power spectrum of the clean speech signal and $\Vert n (t, b) \Vert \in \mathbb{R} _{\geq 0}$ is the power spectrum of the noise signal at each time-frequency bin $(t, b)$.
The ideal binary noise mask $\textrm{IBM}_{\text{N}} (t,b) \in \{0, 1\}$ is also calculated in a similar way.

Given a sequence of $T$-length noisy speech magnitude spectra $Y = (\{\Vert y (t, b) \Vert \} _{b=1} ^B | t =1, \cdots, T)$, the BLSTM network predicts the speech mask $w _{\text{X}} (t, b) \in [0, 1]$ and noise mask $w _{\text{N}} (t, b) \in [0, 1]$ at each time-frequency bin $(t, b)$, as follows:
\begin{equation}
 w _{v} (t, b) = \sigma(\text{Lin} _{v} (\text{BLSTM} (Y))), \  \textrm{where} \  v \in \{\text{X}, \text{N}\},
\end{equation}
where $\sigma(\cdot)$, $\text{Lin} _{v}(\cdot)$, and $\text{BLSTM}(\cdot)$ are sigmoid activation, linear, and output of the masking network given in Table \ref{tab:para}, respectively.

The binary cross entropy is used as the loss function to train the network. 
The overall loss function is defined by combining the speech and noise binary loss functions ($\text{loss}_{\text{X}}$ and $\text{loss}_{\text{N}}$) as follows:
\begin{align}
\label{eq:original_loss}
\text{loss} & =  \text{loss}_{\text{X}} + \text{loss}_{\text{N}},\\
& \triangleq \frac{1}{T*B}\sum _{t, b} \sum _{v \in \{\text{X}, \text{N}\}} \text{CE} (\text{IBM}_{v} (t, b), w _{v} (t, b)),
\end{align}
where $\text{CE}(a, a')$ is the binary cross entropy defined as follows:
\begin{align}
 \label{eq:binary_CE}
 \text{CE}(a, a') & \triangleq a \log a' + (1 - a) \log (1 - a') \\
 & = 
\begin{cases}
 a \log a' & \quad a = 0 \nonumber \\
 (1 - a) \log (1 - a')  & \quad a = 1 \\
\end{cases}.
\end{align}
Note that since $a \in \{0, 1\}$, either first or second term in the right hand side of Eq.~\eqref{eq:binary_CE} becomes zero.
This type of target is particular called as a {\it hard label} in the student-teacher learning \cite{nips_st} or knowledge distillation \cite{hinton2015distilling} context.

When we apply the estimated speech mask $w _{\text{X}} (t, b)$ to the original signal $y(t, b)$, we can perform single-channel speech enhancement as follows:
\begin{equation}
 x ^{\text{(se)}} (t, b) = w _{\text{X}} (t, b) y(t, b),
\end{equation}
where $x ^{\text{(se)}} (t, b)$ is an enhanced signal (the superscript (se) denotes single-channel enhancement).
Although the single channel enhancement only requires the prediction of the speech mask, the noise mask is used to obtain the noise PSD matrix in a multichannel scenario, which will be discussed in the next section.

\begin{table}[tb]
  \caption{Masking network architecture}
  \label{tab:para}
  \centering
  \begin{tabular}{ c | c | c }  
    \toprule
    \textbf{Layer} & \textbf{Activation} & \textbf{Dimension}\\
    \midrule
    Input & - & 513 \\
    BLSTM & Tanh & 256 \\
    Feedforward 1 & ReLU & 513\\ 
    Feedforward 2 & clipped ReLU & 513\\ 
    \bottomrule
  \end{tabular}
\end{table}

\subsection{Mask-based beamformer}
This section extends to deal with a multichannel signal with $M$ as a number of channels. 
When the single channel mask estimation in the previous section is applied to the multichannel signal, the corresponding speech mask $w _{m, \text{X}} (t, b)$ and noise mask $w _{m, \text{N}} (t, b)$ for each channel $m$ can be obtained.
With these multichannel masks, the following time-frequency mask is obtained from a median operation: 
\begin{equation}
 \bar{w} _{v} (t, b) = \text{Median} (\{w _{m, v} (t, b)\} _{m=1} ^M) \textrm{ where } v \in \{\text{X}, \text{N}\}. 
\end{equation}
With the speech and noise masks $\bar{w} _{\text{X}} (t, b)$ and $\bar{w} _{\text{N}} (t, b)$, the PSD matrices, which are 
$\mathbf{\Phi}_{\text{X}}(b) \in \mathbb{C} ^{M\times M}$ of speech at frequency bin $b$, and $\mathbf{\Phi}_{\text{N}} (b) \in \mathbb{C} ^{M \times M}$ of noise at frequency bin $b$, can be estimated as follows:
\begin{equation}
\mathbf{\Phi}_{v} (b) = \sum\limits_{t=1}^T \bar{w}_v(t,b)\mathbf{y}(t, b)\mathbf{y}(t, b)^\textrm{H} \quad \textrm{where} \quad v \in \{\text{X}, \text{N}\} 
\end{equation}
where $\mathbf{y}(t, b) \in \mathbb{C} ^{M}$ is the $M$-dimensional complex spectral value at time (frame) $t$ and frequency bin $b$.

From these PSD matrices, the $M$-dimensional complex beamforming filter $\mathbf{f} (b) \in \mathbb{C} ^M$ can be estimated.
This paper adopts the GEV beamformer, which is obtained by maximizing the expected SNR with respect to $\mathbf{f} (b)$ for each frequency bin as given by the equation below:
\begin{equation}
\mathbf{f}_{\textrm{GEV}} (b) = \argmax_{\mathbf{f}  (b)} \frac{\mathbf{f}^\textrm{H} (b)\mathbf{\Phi}_{\text{X}} (b) \mathbf{f} (b)}{\mathbf{f}^\textrm{H}  (b) \mathbf{\Phi}_{\text{N}} (b) \mathbf{f} (b)}.
\end{equation}
This optimization is equivalent to solving the following eigenvalue problem:
\begin{equation}
(\mathbf{\Phi}_{\text{N}} (b) ) ^{-1} \mathbf{\Phi}_{\text{X}} (b) \mathbf{f}  (b) = \lambda \mathbf{f}  (b),
\end{equation}
where $\mathbf{f}  (b)$ is the eigenvector and $\lambda$ is the corresponding eigenvalue. 

Once we obtain the beamforming filter $\mathbf{f}_{\textrm{GEV}} (b)$, we can perform multichannel speech enhancement as follows:
\begin{equation}
 x ^{\text{(me)}} (t, b) = \mathbf{f}_{\textrm{GEV}}^\textrm{H} (b)  \mathbf{y}(t, b),
\end{equation}
where $x ^{\text{(me)}} (t, b)$ is an enhanced signal  (the superscript (me) denotes multichannel enhancement).

In general, $x ^{\text{(me)}} (t, b)$ can be well-denoised by making use of spacial information and beamforming operation, compared to the single-channel enhanced signal $x ^{\text{(se)}} (t, b)$ introduced in Section \ref{sec:sing}.
This paper proposes to use these single- and multi-channel speech enhancement properties, and designs a new objective function for single-channel enhancement, by using the mask obtained by multichannel enhancement as a better {\it soft label}.

\begin{table*}[tbh]
  \caption{WER of HMM-GMM ASR System}
  \label{tab:1ch}
  \centering
  \scalebox{0.90}{
  \begin{tabular}{@{\makebox[3em][r]{\rownumber\space}} | c c c c | c | c  | c c | c c }
    \toprule
 \multicolumn{4}{c}{\textbf{Parameters}} & \multicolumn{2}{c}{} & \multicolumn{2}{c}{\textbf{WER Dev (\%)}} & \multicolumn{2}{c}{\textbf{WER Test (\%)}} \\
   $\lambda_1$ & $\lambda_2$ & $\lambda_3$ & epoch & Train data (ASR) & BLSTM Mask & real & simu & real & simu 
   	
    \gdef\rownumber{\stepcounter{magicrownumbers}\arabic{magicrownumbers}} \\
    \midrule
    - & - & - & - & all 6ch noisy & - & \textbf{21.40} &  \textbf{23.22} & \textbf{35.63} & \textbf{31.98} \\
    - & - & - & - & all 6ch noisy & Baseline & 28.99 &  28.05 & 40.98 & 35.50 \\
    - & - & - & 7 & all 6ch noisy & Teacher & 24.91 &  26.00 & 40.26 & 35.73 \\
    \midrule
    1/3  & 1/3 & 1/3 & 6 & all 6ch noisy & Student & 25.95 &  24.66 & 35.50 & 29.98 \\
    0.25  & 0.25 & 0.5 & 12 & all 6ch noisy & Student& 26.56 &  26.19 & 36.33 & 31.36 \\
    0.50  & 0.25 & 0.25 & 5 & all 6ch noisy & Student& 25.82 &  25.17 & 35.86 & 30.17 \\
    0.45  & 0.05 & 0.50 & 3 & all 6ch noisy & Student& 25.54 &  24.77 & 34.78 & 29.78 \\
    0.35  & 0.15 & 0.50 & 3 & all 6ch noisy & Student& \textbf{23.34} &  \textbf{23.11} & 33.11 & \textbf{28.30} \\
    0.05  & 0.45 & 0.50 & 3 & all 6ch noisy & Student& 24.07 &  24.01 & 34.88 & 30.00 \\
    0.15  & 0.35 & 0.50 & 3 & all 6ch noisy & Student& 25.01 &  24.66 & 35.66 & 30.10 \\ 
    0.35  & 0.15 & 0.50 & 3 & all 6ch noisy & Student with real & 23.42 &  23.55 & \textbf{32.64} & 28.88 \\
    \midrule
 - & - & - & - & \shortstack{all 6ch noisy +\\ 5th ch enhanced data from baseline} & Baseline & 22.07 &  23.37 & 34.02 & 30.41 \\
0.35  & 0.15 & 0.50 & 3 & \shortstack{all 6ch noisy +\\ 5th ch enhanced data from baseline} & Student & \textbf{19.78} &  \textbf{20.76} & 30.66 & \textbf{26.60} \\
0.35  & 0.15 & 0.50 & 3 & \shortstack{all 6ch noisy +\\ 5th ch enhanced data from baseline} & Student with real& 19.79 &  20.85 & \textbf{29.80} & 26.66 \\
    
    \bottomrule
  \end{tabular}
  }
\end{table*}

\begin{figure}[t]
  \centering
  \includegraphics[width=\linewidth]{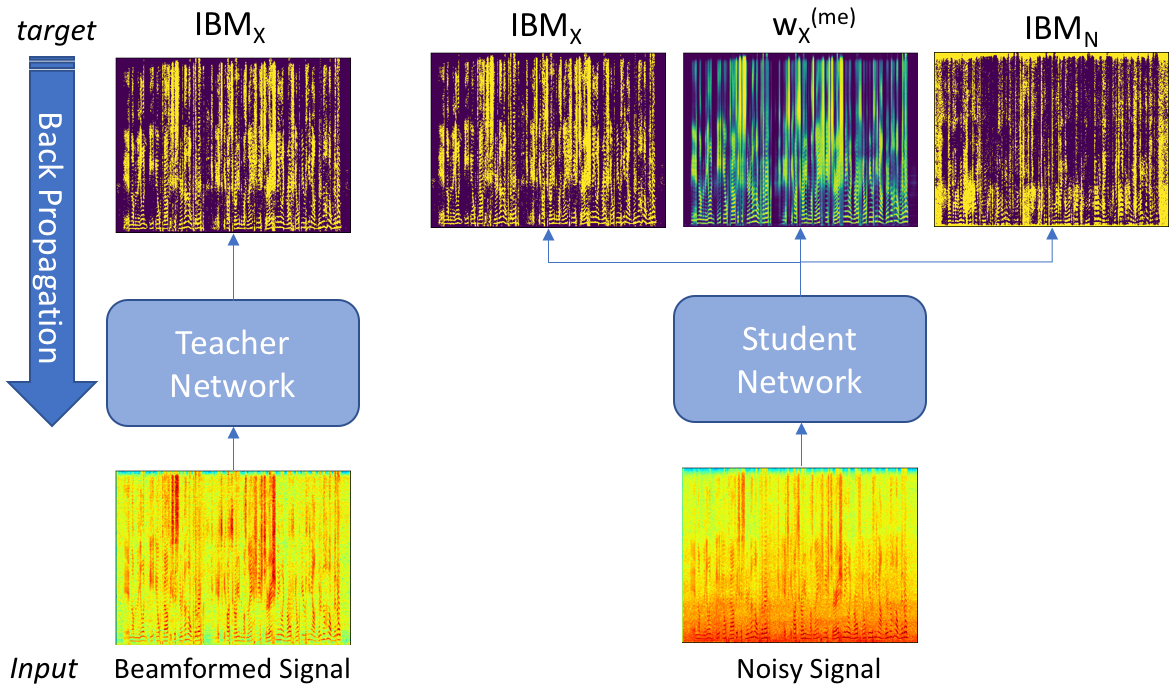}
  \caption{Student-Teacher Model}
  \label{fig:st_model}
\end{figure}

\section{Student-teacher model}
This section explains in detail the proposed single-channel enhancement technique by using a student-teacher model.
\subsection{Teacher model}
Firstly, the beamformed signal $x ^{\text{(me)}} (t, b)$ is given as input to the teacher model. 
The architecture of the network is the same as the one explained in Section \ref{sec:sing} except that the output is only the speech mask $w _{\text{X}} ^{\text(me)} (t, b)$ and there is no noise mask. 
Hence the binary loss function for a teacher network is 
\begin{equation}
\text{loss} = \frac{1}{T*B}\sum _{t, b} \text{CE} (\text{IBM}_{\text{X}} (t, b), w _{\text{X}} ^{\text(me)} (t, b))).
\end{equation}
Since the beamformed signal is already well-denoised, the above teacher network provides a high-quality speech mask.
Alternatively, the original clean signal $x (t, b)$ can be used as the input instead of the beamformed signal to train the teacher model.
However, the problem of estimating a speech mask given a clean signal is too trivial, and the network is not well-trained in such a trivial condition.



\subsection{Student model}
With the speech mask $w _{\text{X}} ^{\text(me)} (t, b)$ obtained by the above teacher network, we can additionally consider the following student-teacher loss function:
\begin{equation}
\label{eq:stloss}
\text{loss} _{\text{st}} =  \frac{1}{T*B}\sum _{t, b} \text{CE} (w _{\text{X}} ^{\text(me)} (t, b)), w _{\text{X}} ^{\text(se)} (t, b))).
\end{equation}
Compared with the binary cross entropy case in Eq. \eqref{eq:binary_CE}, the target $w _{\text{X}} ^{\text(me)} (t, b)) \in [0, 1]$ is a bounded continuous value, and provides richer supervisions to the student network.

The final loss function is expressed with the student-teacher loss and the original loss (Eq.~\eqref{eq:original_loss}) as:
\begin{equation}
\label{eq:combineloss}
\text{loss} = \lambda_1 \text{loss} _{\text{st}} + 
\lambda_2 \text{loss} _{\text{X}} + \lambda_3 \text{loss} _{\text{N}}
\end{equation}
where $\lambda_1$, $\lambda_2$, and $\lambda_3$ are linear interpolation weights. 
The student model is trained with this loss function, which expects to learn the multichannel enhancement ability within a single-channel enhancement framework. Figure \ref{fig:st_model} illustrates the proposed student-teacher training paradigm. 

\subsection{Student model with real data}
\label{sec:with_real}
All the above formulations are discussed given that we have parallel clean speech ($x$) and noise ($n$), which are not usually obtained in the real recording.
However, the proposed network can obtain the soft target of the real data by simply applying multichannel speech enhancement for the real data.
In this scenario, it has only one student-teacher loss term $\text{loss} _{\text{st}}$ (Eq.~\eqref{eq:stloss}), which exactly mimics the teacher network.
With this setup, the loss function can be obtained by switching the student-teacher loss (Eq.~\eqref{eq:stloss}) and the combined loss (Eq.~\eqref{eq:combineloss}) for real and simulation data as follows:
\begin{equation}
\text{loss} = 
\begin{cases}
\text{loss} _{\text{st}} & \text{ for real}\\
\lambda_1 \text{loss} _{\text{st}} + 
\lambda_2 \text{loss} _{\text{X}} + \lambda_3 \text{loss} _{\text{N}} & \text{ for simulation}
\end{cases}
\end{equation}
This is our final loss function, which fully makes use of the benefit of the student-teacher learning paradigm in speech enhancement.

\section{Experiments}
To verify the effectiveness of the proposed single-channel experiments, the 1 channel track in the CHiME-4 challenge \cite{vincent2017analysis} was used.

\begin{table*}[ht]
  \caption{Speech Enhancement Scores.}
  \label{tab:se_scores}
  \centering
  \scalebox{0.93}{
  \begin{tabular}{ c c c c | c  | c c c c | c c c c }
    \toprule
    \multicolumn{4}{c}{\textbf{Parameters}} & \multicolumn{1}{c}{} & \multicolumn{4}{c}{\textbf{Dev (Simu)}} & \multicolumn{4}{c}{\textbf{Test (Simu)}} \\
   	$\lambda_1$ & $\lambda_2$ & $\lambda_3$ & epoch & BLSTM Mask & PESQ & STOI & eSTOI & SDR & PESQ & STOI & eSTOI & SDR \\
   	\midrule
    - & - & - & - & - & 2.01 & 0.82 & 0.61 & 3.92 & 1.98 & 0.81 & 0.60 & 4.95 \\
    - & - & - & - & Baseline & 2.52 & 0.88 & 0.73 & 9.26 & 2.46 & 0.87 & 0.71 & 10.76 \\
    - & - & - & 7 & Teacher & 2.36 & 0.86 & 0.68 & 8.14 & 2.33 & 0.85 & 0.67 & 9.51 \\
    1/3  & 1/3 & 1/3 & 6& Student & 2.54 & 0.88 & 0.73 & 9.13 & 2.49 & 0.87 & 0.71 & 10.64 \\
    0.50  & 0.25 & 0.25 & 5 & Student & 2.52 & 0.88 & 0.73 & 9.00 & 2.47 & 0.87 & 0.71 & 10.44 \\
    0.35  & 0.15 & 0.50 & 3 & Student & N/A & 0.88 & 0.72 & 8.93 & N/A & 0.87 & 0.70 & 10.35 \\
      \bottomrule
  \end{tabular} 
  }
\end{table*}

\label{sec:asr_exp}

\begin{figure}[t]
  \centering
  \includegraphics[width=\linewidth]{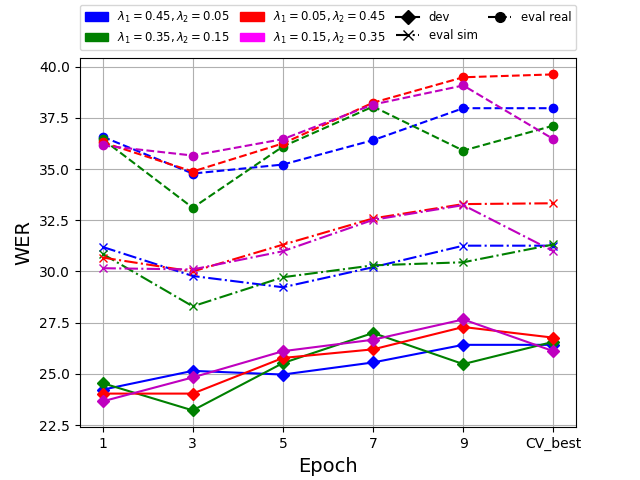}
  \caption{WER vs epoch for different parameter combinations}
  \label{fig:wer}
\end{figure}

\subsection{Speech recognition}
The Kaldi speech recognition toolkit \cite{povey2011kaldi} was used for ASR experiments. The GMM baseline provided by the CHiME-4 challenge \cite{vincent2017analysis} was used for comparing different mask models.
The mask network was implemented using the Chainer neural network toolkit \cite{chainer_learningsys2015}. The cross-validation loss was calculated using the development data after every epoch during the training of the teacher network. 
The training was stopped when the loss did not decrease anymore after 5 epochs of patience and the model with the least cross-validation loss was chosen. 
The 5th channel data was used to estimate the target masks for the beamformed data. 
The student network was trained in a batch mode where a minibatch comprised the frames of all 6 channels of one utterance.

To compare the performance of different masking models, we first prepared an ASR baseline that would be neutral to the enhancement method by using only the noisy data of all 6 channels as the training data. 
We performed enhancement on the development and evaluation sets of the official 1-channel track CHiME-4 dataset, with different models and decoded their enhanced speech with the above ASR systems.
The results of these ASR experiments are shown in Table \ref{tab:1ch} (rows 1--3).
The performance of the teacher model was better than that of the original BLSTM mask as expected, although both degraded the performance from the non-enhanced noisy speech.

The second experiments focused on our proposed student models (rows 4--11 in Table \ref{tab:1ch}), and we found that most of the different configurations of student models performed better than that of the teacher model.
Note that in this experiment, we fixed the number of epochs for several models as 3.
This is because we additionally investigated the dependency between the validation loss and WER, as shown in Figure \ref{fig:wer}, and we found that it did not necessarily give us the best results for the best validation score with different hyper-parameter $\lambda$, as introduced in Eq.~\eqref{eq:combineloss}.
We empirically found that 3 epochs in most of the cases seemed to give good performance. 
This result indicates that choosing the epoch based on the WER of the development data seems to be a better criterion than the enhancement-driven cross-validation loss (related discussions will be shown in the next section). 
With this setup (3 epochs), we found that the parameter choice of $\lambda_1 = 0.35$, $\lambda_2 = 0.15$ and $\lambda_3 = 0.5$ gave the best performance amongst the different values we tried. 
This is an interesting finding, since this result indicates that soft targets obtained by the teacher model would effect the performance more than that of supervised targets.
We also trained the student model with additional real data, as discussed in Section \ref{sec:with_real} (row 11 in Table \ref{tab:1ch}), and this shows the improvement for the real test set owing to the inclusion of the real training data, which is a desired property in real environments.

When the 5th channel of the training data was also enhanced in the same way as the development and evaluation data using the baseline BLSTM mask and included as part of the ASR training (row 12), the performance was slightly better than using the original noisy data for the evaluation data but it was slightly worse for the development data. When the development and evaluation data was enhanced using our best student models (rows 13 and 14), the performance improved significantly compared to using the original noisy data in the all conditions. Also, we observed the similar improvement for the real test set again when we trained the student model with real data.

\subsection{Speech enhancement scores}
The four different scores described in Section \ref{sec:intro} - PESQ, STOI, eSTOI and SDR were computed for several enhancement methods, as shown in Table \ref{tab:se_scores}. 
The 5th channel clean signal from the 6ch track convolved with room impulse response was used as the reference signal. 
All the different  masking models gave significantly better scores in all four metrics compared to using the noisy data without any enhancement although we did not observe any considerable difference in the scores  amongst the models.
In addition, by comparing Tables \ref{fig:wer} and \ref{tab:se_scores}, we did not observe clear correlations between the speech enhancement scores and WERs, which also suggests that we need some careful investigation on the objective function of BLSTM-based enhancement for the ASR purpose.

\section{Conclusion}
We proposed a new training paradigm for mask estimation using BLSTMs based on student-teacher training for single channel speech enhancement. 
We showed that the proposed student-teacher technique improved the ASR performance from the original noisy speech and the enhanced speech obtained by conventional BLSTM masking. 
Our future work is to evaluate our speech enhancement techniques with a strong ASR backend including a time delay neural network (TDNN) \cite{waibel1990phoneme,peddinti2015time} with the lattice-free version of the maximum mutual information (LF-MMI) \cite{povey2016purely}.


\bibliographystyle{IEEEtran}
\bibliography{mybib}


\end{document}